\documentclass[11pt,
epsfig]{article}
\usepackage{epsfig}
\usepackage{graphicx}

\newcommand{\be}{\begin{equation}}
\newcommand{\ee}{\end{equation}}
\newcommand{\een}{\end{subequations}}
\newcommand{\ben}{\begin{subequations}}
\newcommand{\beq}{\begin{eqnarray}}
\newcommand{\eeq}{\end{eqnarray}}
\newcommand{\nn}{\nonumber}

\begin{document}
\centerline{ \hfill hep-th/0210091 }
\vspace{3.5cm}

\centerline{\Large\bf Scaling Violations in Yang-Mills
Theories and Strings in  $\rm{AdS_{5}}$
} \vspace{2.cm}

\centerline{
 {\large \bf Minos Axenides$^{a}$}\footnote {e-mail
address : axenides@inp.demokritos.gr }{\large ,} {\large \bf Emmanuel
Floratos$^{a,b}$}\footnote {e-mail address :
floratos@inp.demokritos.gr }{\large ,} {\large and} {\large\bf Alex
Kehagias $^{a,c}$}\footnote {e-mail address :
kehagias@mail.cern.ch} } \vspace{0.3cm} \centerline{\em a)
Institute of Nuclear Physics, N.C.S.R. Demokritos, GR-15310
Athens, Greece} \vspace{0.2cm} \centerline{\em b) Nuclear and
Particle Physics Sector, University of Athens, GR-15771 Athens,
Greece} \vspace{0.2cm} \centerline{\em c) Department of Physics,
NTUA, GR-15773, Zografou, Athens, Greece} \vspace{3.cm}

\centerline{\large\bf Abstract}
\begin{quote}\large\indent
String solitons in $AdS_{5}$ contain information of ${\cal N}=4$
SUSY Yang-Mills theories on the boundary. Recent proposals for
rotating string solitons reproduce the spectrum for anomalous
dimensions of Wilson operators for the boundary theory. There are
possible extensions of this duality for lower supersymmetric and
even for non-supesymmetric Yang-Mills theories. We explicitly
demonstrate that the supersymmetric  anomalous dimensions of
Wilson operators in ${\cal N}=0,1$ Yang-Mills theories behave, for
large spin $J$, at the two-loop level in perturbation theory, like
$\log J$. We compile the analytic one- and two-loop results for
the ${\cal N}=0$ case which is known in the literature, as well as
for the ${\cal N}=1$ case which seems to be missing.

\end{quote}
\vspace{1cm}

\newpage

\section{Introduction}






Recent developments in understanding the duality between gravity
and gauge interactions points to a new synthesis of ideas about the
role of string theory
for the infrared behavior of both supesymmetric and non-supersymmetric
Yang-Mills theories. These developments came from three
different directions.

Firstly the $AdS/CFT$ correspondence was extended to new exact
string backgrounds, the so called pp-waves, which are unique and
universal limits (Penrose limit) of every  space-time which admits
null geodesics \cite{papado,Maldac}. In these backgrounds it is
frequently possible to solve exactly the quantum spectrum of
strings while the correspondence with the conformal field theories
on the boundary remains intact. One hopes to get nontrivial
information about interesting boundary theories by using more
detailed information from the string side\cite{FT}-\cite{GMR}.
Penrose limits of backgrounds with various number of supersymmetries
constructed as orbifolds and orientifolds have been discussed in
\cite{Alishahiha:2002ev,Floratos:2002uh,Maldacena:2002fy}.

A second development came from the realization that the free
string Hamiltonian in $AdS_{5}$ background describes a string with
spacetime dependent tension, whereby it develops hard components
with field theory point-like behavior. The hard component of the
$AdS_{5}$ strings appears in the energy scaling behavior of the
production cross-sections of the process $2\rightarrow n$ strings.
This process has been calculated in \cite{Pol2} and found to be
similar to the hard scattering processes of QCD. In the language
of the old parton model the string in flat space-time is very
soft. If viewed as a hadron its average radius  diverges
logarithmically with the number of partons(wea partons).
Interestingly in a $AdS_{5}$ background its average radius is
finite and calculable around a fixed distance from the boundary of
$AdS_{5}$\cite{Pol1}. More recently the dual picture for deep
inelastic processes has also been studied\cite{Pol3}.

The third development concerns  the duality between space-time
geometry and gauge interactions. In \cite{Polyak2}, an explicit
classical string soliton solution has been found in $AdS_{5}$
which represents a collapsed closed string in the form of a rod
rotating with constant angular velocity in the equator of $S^{3}$
of $AdS_{5}$. By considering the deviations from flat space-time
for the energy-angular momentum relations for large spin,
logarithmic corrections were found similar to the large spin
behavior of the anomalous dimensions of Wilson operators for the
${\cal{N}} =4$ supersymmetric Yang-Mills theories on the boundary.
It is interesting to note that similar behavior was found for
rotating long strings in the $AdS_{5}$ black hole
background\cite{Petkou} while rotating strings exhibiting
confinement as well as finite-size effects have been studied in
\cite{Armoni:2002fr}. It should be stressed here that for the GKP
solution which describes the rotating string in $AdS_5$, the
internal space does not enter anywhere. Thus, the superstring
background can be the standard maximally supersymmetric
$AdS_5\times S^5$, as well as any of the available $N=2$
supergravity backgrounds of the form $AdS_5\times X^5$ described
in \cite{Keha}. It can even be  a non-supersymmetric background
with an $AdS_5$ factor. This seems to indicate that the the large
spin behavior of the anomalous dimensions of Wilson operators in
Yang-Mills theories is the same (up to possible coefficient
differences) for both supersymmetric and non-supersymmetric
theories.

In this work we confirm that both ${\cal N}=0$ and ${\cal N}=1$
gauge theories exhibit identical large spin behavior. In
particular, we reconsider in detail the existing calculations for
anomalous dimensions of Wilson operators in ${\cal N}=0$
Yang-Mills theories up to two loops.  Moreover for the case of
${\cal N}=1$ we calculate explicitly the two loop anomalous
dimensions from the known results of the non-supersymmetric
case\cite{Flo3}. We identify the numerical coefficients in front
of the confirmed logarithmic behaviour for large spin. In sect.$2$
we recall the GKP rotating strings \cite{Polyak2} in $AdS_{5}$ in
the limit of large spin (long strings). In sect. $3$ we review the
formalism of operator product expansion for deep inelastic
scattering and we set up our notation. In sect. $4$ we exhibit the
known analytic results of the two loop anomalous dimensions for
the ${\cal N}=0$ pure Yang-Mills theories. In sect. $5$ we present
the analytic results ${\cal N}=1 $ results and we give the
asymptotic behavior of both supersymmetric and non-supersymmetric
cases in the limit of large spin.

\section {$AdS_{5}$ Rotating Strings in the Large Spin Limit}

One of the emerging scenarios, due to Polyakov, provides a further
extension to the $AdS/CFT$ correspondence between the $AdS_{5}$
supergravity and the boundary theory of ${\cal N}=4$
supersymmetric Yang-Mills which possibly may reach out all the way
to the non-supersymmetric regime. According to it the $AdS_{5}$
space-time which appears as a solution of the quantum
 non-critical string theory in four dimensions provides a dual description
of pointlike field  theories. In such a description the fifth
dimension plays the role of a
 non-critical Liouville field\cite{Polyak2,Polyak1}.
The non-critical string represents the dynamics of gauge field
strength lines and it has to live in $AdS_{5}$. The $AdS_5$ radius
$R$ satisfies the relation $ R^{4}=\lambda \alpha'^{2}$ where
$\lambda=(g^{2}_{YM} N)/4\pi$ is the 't Hooft coupling. Weak
coupling string interactions correspond to strong gauge
interactions on the boundary. As a result,
 in order to explore
the weak gauge coupling regime or equivalently the high energy
behavior of the boundary gauge theory, one has to study
nonperturbative classical string theory ( large N behavior) along
with its quantum corrections. The GKP rotating string soliton in
$AdS_{5}$ offers an intriguing playground for developing and
understanding these ideas. In particular, it may provide a useful
tool in exploring
 the transition
region between weak and strong gauge couplings or small and
big space-time
curvatures.

In our present paper we make explicit quantitative comparison
between string soliton behavior for large 't Hooft coupling and
the analytic results for two loop anomalous dimensions of
${\cal{N}}=0,1$ Yang-Mills theories in the same limit.
 We begin with the description of the Polyakov string soliton. We follow the
parametrization of the global $AdS_{5}$ metric
\beq
 ds^{2}= R^{2}( -dt^{2}\cosh^{2}\rho +d\rho^{2}+\sinh^{2}\rho
d\Omega_{3}^{2}) \eeq The string soliton rotates at the equator of
$S^{3}$ and the azimuthal angle depends linearly on time \beq \phi
= \omega t \eeq By choosing the timelike gauge $ t=\tau$ and by
assuming that the radial coordinate $\rho$ is only a function of
$\sigma$ we obtain the Nambu-Gotto Lagrangian \beq L=
-4\frac{R^{2}}{2\pi \alpha'}\int_{0}^{\rho_{0}} d\rho\;\sqrt{
\cosh^{2}\rho - \dot{\phi}^{2}\sinh^{2}\rho} \eeq The maximun
radial distance is $\rho_{0}$ which is determined by the speed of
light \beq \coth^{2}\rho_{0}=\omega^{2} \eeq The reparametrization
constraints give the equation \beq \rho'^{2}= e^{2}(\cosh^{2}\rho-
\omega^{2}\sinh^{2}\rho) \eeq where $\rho^{\prime}=
\frac{d\rho}{d\sigma}$ and $e$ is adjusted so that $\sigma$ has a
period $2\pi$. The space-time energy E and spin J of the rotating
string are then given by

 \beq E&=&
\frac{R^{2}}{2\pi\alpha^{\prime}}e \int_{0}^{2\pi}d\sigma\;
\cosh^{2}\rho\nn \\ S&=&\frac{R^{2}}{2\pi\alpha^{\prime}}e\omega
\int_{0}^{2\pi}d\sigma \;\sinh^{2}\rho \eeq These expressions
determine $E/\sqrt{\lambda}$ and $J/\sqrt {\lambda}$ as functions
of $\omega$.

We are interested here in  the classical limit $J \gg
\sqrt{\lambda}$ which corresponds to $\omega$ approaching one from
above $\omega= 1 +\eta$ with $0<\eta\ll 1$. In this limit the end
of the string approaches the boundary of the $AdS_{5}$
$\rho_{0}\rightarrow \frac{1}{2}\log(1/\eta)$. By expanding the
energy and spin in terms of $\eta$ we obtain

\beq
E-J\sim \sqrt{\lambda}\;\;\log\frac{J}{\sqrt{\lambda}}+\cdots
\eeq

On the gauge theory side, $E-J$, i.e,  the dimension minus
the spin of some composite operator is the twist of the operator.
In the case of the leading trajectory
(solid rotating string), the leading contributions come from
 twist-two operators.
The deviation from the flat space-time is reminiscent of the
behavior of the anomalous dimensions of the twist-two Wilson
operators of deep inelastic scattering for large spin. In a sense,
the anti-deSitter backround forces the string to develop hard
partonic component reminiscent of QCD \cite{Pol2,Pol1}. In
\cite{Polyak2} a concrete conjecture about the large J behavior of
$E-J$ is made to the effect that the leading term for arbitrary
$\lambda$ will be \beq E-J = f(\lambda)\;\log J \eeq In \cite{FT},
the $1-$loop corrections were computed
 \beq
f(\lambda)= \frac{1}{\pi}\sqrt{\lambda} - \frac{3}{4\pi}\log 2
\eeq
and no higher powers of $\log J$ corrections were found.
These calculations hold for large $\lambda$ but with $J\gg
\lambda$. One hopes that it will be possible to analytically
continue these results for small values of $\lambda$ in order to
make contact with the perturbative results for the Yang-Mills
theories on the boundary. This is the subject of the recent work
by Tseytlin et.al in \cite{FT}.


\setcounter{equation}{0}

\section{Operator Product Expansion in Deep-Inelastic Scatering}

In the experiments of deep-inelastic scattering of leptons on
nucleons and in
the one electroweak gaugeboson exchange approximation, one measures the
differential  cross sections of the emerging lepton with specific final
 four-momentum for unpolarized scattering\cite{Gross,GW}
\beq
{d\sigma\over d^4p_l}= l^{\mu\nu}\cdot W_{\mu\nu}
\eeq
with
\beq
l^{\mu\nu}=4(p_i^\mu p_f^\nu+p_i^\nu p_f^\mu-g_{\mu\nu}p_i\cdot p_f)
\eeq
and
\beq
W_{\mu\nu}={1\over 8\pi} \sum_{\mbox{spins}}\int d^4 x e^{iq\cdot x}
<p,s|[J^\mu(x),J^\nu(0)]|p,s>
\eeq
In the Bjorken limit $Q^2=-q^2\to \infty$ and $x_B={Q^2\over 2p\cdot q}$ fixed
where $q=(p_f-p_i)$ is the momentum transfer and $p$ is the nucleus momentum
the Fourier transform above is dominated by  light-cone distances. The
commutator can be calculated as the imaginary part of the time-ordered product
of the electroweak currents $J^\mu$
\beq
W_{\mu\nu}&=&{1\over 2\pi} {\rm Im} T_{\mu\nu}(p,q)\nonumber \\
&=&e_{\mu\nu}{1\over 2 x_B}
F_L(x_B,Q^2)+d_{\mu\nu}{1\over 2x_B}F_2(x_B,Q^2)+i\epsilon_{\mu\nu\alpha\beta}
{p^\alpha q^\beta\over p\cdot q}F_3(x_B,q^2)
\eeq
where
\beq
T_{\mu\nu}=i\int d^4x e^{iq\cdot x}  <p|T(J^\mu(x)J^\nu(0))|p>
\eeq
and $F_L,F_2,F_3$ are the structure functions which are measured by
the experiments ($F_3$ is relevant for neutrinos).
The right-hand side is dominated by the light-cone $x^2\to 0$
and   can be expanded
in terms  of composite operators multiplied by
coefficient functions
\beq
T_{\mu\nu}=\sum_{N,i}\left({1\over 2 x_B}\right)^N\left[
e_{\mu\nu}C^N_{L,i}({Q^2\over \mu^2},\alpha_s)+d_{\mu\nu}
C^N_{2,i}({Q^2\over \mu^2},\alpha_s)+i\epsilon_{\mu\nu\alpha\beta}
{p^\alpha q^\beta\over p\cdot q} C^N_{3,i}({Q^2\over \mu^2},\alpha_s)\right]
A^N_{i}({p^2\over \mu^2}) \label{Tmn}
\eeq
where
\beq
e_{\mu\nu}&=&g_{\mu\nu}-{q_\mu q_\nu\over q^2}\nonumber \\
d_{\mu\nu}&=& -g_{\mu\nu}-p_\mu p_\nu {4 x_B^2\over q^2}-(p_\mu q_\nu
+p_\nu q_\mu){2 x_B\over q^2}
\eeq
$\alpha_s={g^2_{YM}\over 16 \pi^2}$ and
 $A^N_i$ are the matrix elements of the dominant  operators
$O_{i}^N$ ($i={\rm NS,quark,gluons})$ beteween the nucleon state.
The dominant terms in the light-cone expansion come from lowest
twist (dimension minus spin) operators which can be constructed
from the Yang Mills theory with fermions. They are the flavour
non-singlet (valence quarks) and the singlet ones (sea-quarks and
gluons) \beq O_{NS,\alpha}^n={i^{n-1}\over 2
n!}\left(\bar{q}\lambda^\alpha\gamma^{\mu_1} D^{\mu_2}\cdots
D^{\mu_n}(1\pm \gamma^5)q+\mbox{permutations-traces}\right) \eeq
where $\alpha$ is a flavour index and the chiral projector appears
in the scattering of neutrinos and anti-neutrinos while in the
scattering of electrons or positrons is missing. The singlet
operators are \beq O_{q}^n&=& {i^{n-1}\over
n!}\left(\bar{q}\gamma^{\mu_1} D^{\mu_2}\cdots
D^{\mu_n}q+\mbox{permutations-traces}\right)\nonumber \\
O_{g}^n&=& {i^{n-2}\over  2n!}{\rm Tr}\left(F^{\lambda\mu_1}
D^{\mu_2}\cdots
D^{\mu_{n-1}}{F_\lambda}^{\mu_n}+\mbox{permutations-traces}
\right) \eeq These operators are the twist $2=(n+2)-n$ ones which
are dominating in the light-cone expansion. From eq.(\ref{Tmn})
taking moments over the Bjorken variable $x_B$ we project out the
spin $N$ term \beq \int_0^1 dx\,  x^{N-k}
F_i(x,Q^2)=\sum_{i=NS,q,g} C^N_{i,j}({Q^2\over \mu^2},\alpha_s)
A^N_j({p^2\over \mu^2}) \label{mom} \eeq where $k=2$ for $F_2,F_L$
and $k=1$ for $F_3$. The importance of the moment equations
(\ref{mom}) comes from the renormalization properties of the
Wilson operators \beq <p|O_i^N|p>=\left(p^{\mu_1}\cdots
p^{\mu_N}-{\rm traces}\right) A^N_i({p^2\over \mu^2}) \eeq The
non-singlet one is multiplicative renormalized \beq O^N_{NS,{\rm
bare}}= Z_{NS}^N O^N_{NS,{\rm ren.}} \eeq while the singlet ones
between physical states  mix through a two-by-two
 matrix renormalization constant
\beq
O^N_{i,{\rm bare}}=Z_{ij}^N O^N_{j,{\rm ren.}}\, , ~~~ i,j=q,g
\eeq
Because of these properties, the moment equations give information about the
$Q^2$ evolution of structure functions. Indeed, since the left-hand side
is renormalization group invariant, the dependance on the
renormalization scale $\mu$ of the right-hand side
should cancel between the coefficient functions and the operator matrix
elements. This gives the renormalization
group equations for the coefficient functions
\beq
\left(\mu^2 {\partial\over \partial
\mu^2}+\beta(\alpha_s){\partial\over \partial
\alpha_s}-\gamma^N_{NS}(\alpha_s)\right) C^N_{NS,i}
({Q^2\over \mu^2},\alpha_s)&=&0\nonumber \\
\sum_k \left[\left(\mu^2 {\partial\over \partial
\mu^2}+\beta(\alpha_s){\partial\over \partial
\alpha_s}\right)\delta_{jk}-\gamma^N_{jk}(\alpha_s)\right] C^N_{i,k}
({Q^2\over \mu^2},\alpha_s)&=&0 \, , ~~~~j,k=q,g
\eeq
The expansion of the $\beta$- and $\gamma$-function in $\alpha_s$
can be calculated from the renormalization constants by looking at the
coefficient of the simple poles in the dimensional regularization
parameter $\epsilon$, $d=4-\epsilon$
\beq
\beta={1\over 2}\alpha_s {\partial\over \partial \alpha_s}Z^{(1)}_{\alpha_s}
\, , ~~~
\gamma^N_{NS}=-\alpha_s{\partial\over \partial \alpha_s}
Z^{(1),N}_{NS}({\alpha_s})\, ,
~~~
\gamma^N_{ij}=-\alpha_s{\partial\over \partial \alpha_s}
Z^{(1),N}_{ij}({\alpha_s})
\eeq
where typicaly
\beq
Z({\alpha_s})=1+{Z^{(1)}({\alpha_s})\over \epsilon}+
{Z^{(2)}({\alpha_s})\over \epsilon^2}+\cdots
\eeq

\setcounter{equation}{0}
\section{The Two Loop Anomalous Dimensions of Wilson Operators:
Generalities and Old Results}

The anomalous dimensions of Wilson operators $O^N_{i}$ at one and
two loop level for any representation of fermions are known for
the following cases. For ${\cal N}=0$ the anomalous dimension for
QCD were calculated firstly at one loop in \cite{Gross,GW} and at
two loop levels in \cite{Flo1}. It has been recalculated in
\cite{CFP} and a discrepancy was found for the two loop
$\gamma_{gg}$ in the coefficient of the Casimir $C^{2}_{A}$. In
\cite{Flo3} through the supersymmetric identities that will be
discussed below the discrepancy was resolved in favor of
\cite{CFP}. This coefficient has been recalculated in \cite{HvN}
and the result was found to be in agreement with that of
\cite{CFP}. The complete two loop QCD results were recalculated
 in \cite{Flo2,Ver}. In \cite{GLY}, the two-loop anomalous
dimensions were written in a very compact form and for
phenomenological reasons the large spin  behavior was studied and
the asymptotic $(log J)$ behavior was singled out.

The ${\cal N}=1$ case was studied in one-loop by \cite{DDT} where
the supersymmetric identity for the singlet anomalous dimensions
was found \beq
\gamma^{(0)}_{qq}(J)+\gamma^{(0)}_{gq}(J)=\gamma^{(0)}_{qg}(J)+
\gamma^{(0)}_{gg}(J) \label{qqgg} \eeq for every $J$.   This
infinite number of relations, is due to the fact  that the
combination of the Wilson operators $O^J_{q}+O^J_{g}$ is
multiplicatively renormalized \cite{Flo3,DDT,Belitsky:2000xk}. For
higher supersymmetries ${\cal{N}}=2,~{\cal{N}}=4$ there are
similar relations involving the scalar operator anomalous
dimensions \cite{KR,Lip}. At the two-loop order one can check that
the anomalous dimension singlet matrix elements obtained above
satisfy the same relations\cite{Flo3}. These infinite number of
relations provide an important check of the calculation and at the
same time they confirm that the dimensional reduction scheme(DR)
is the appropriate one for supersymmetric gauge
theories\cite{Jones}.

 The ${\cal N}=1$ case with quark multiplets (SUSY-QCD)
was studied for phenomenological reasons at the two-loop level in
the approximation of light gluinos and heavy squarks \cite{AKL}.
The contribution of heavy squarks was omitted from the two loop
anomalous dimensions and the $Q^{2}$ evolution of the structure
functions. As a result the ${\cal N}=2$ (quark hypermultiplet in
the adjoint) two-loop anomalous dimensions could not be obtained,
as the ${\cal{N}}=1$ is incomplete. On the other hand the complete
${\cal{N}}=2$ anomalous dimensions were obtained at one-loop level
and additional supersymmetric relations for the singlet case were
found in \cite{KR}.

The ${\cal N}=4$ anomalous dimensions at one loop are also known
and it is claimed that the two loop result can be obtained from
the analytic properties of the DGLAP and BFKL evolution kernel
\cite{Lip}.

For completeness of the presentation and in order to prepare the
ground for the ${\cal N}=1$ case we present below the old known
two loop ${\cal N}=0$ results in the compact form following
\cite{Flo2,GLY}. We expand the anomalous dimensions typically as :

\beq\label{ZERO} \gamma(J)= \gamma^{(0)}(J)
\frac{\alpha_{s}}{4\pi} +
\gamma^{(1)}(J)(\frac{\alpha_{s}}{4\pi})^{2} + \cdots \eeq

 \beq\label{ENA} \gamma^{(1)}_{NS}(J)&=&
C_{F}^{2}\left(\frac{16S_{1}(J)(2J+1)}{J^{2}(J+1)^{2}} +
16\Big{(}2S_{1}(J)-{1\over J(J+1)}\Big{)}
(S_{2}(J)-S_{2}^{\prime}(J))
 + \right.\\& &\left. \mbox{} 24
S_{2}(J)+64\tilde{S}(J)  -8 S_{3}^{\prime}(J) -3 - \frac{8(1+4 J+5
J^2+3 J^3)}{J^{3}(J+1)^{3}}\right)\nn \\& &\mbox{} + C_{A}C_{F}
\left( \frac{536}{9}S_{1}(J)-8\Big{(}2S_{1}(J)
-\frac{1}{J(J+1)}\Big{)}(2S_{2}(J)-S_{2}^{\prime}(J))-
\frac{88}{3}S_{2}(J) -28\tilde{S}(J) \right.\nn
\\& &\left. \mbox{}-\frac{17}{3}-\frac{4}{9}\frac
{(-33+52J+236J^2+151J^3)}{J^{2}(J+1)^{3}}\right)\nn \\&
&\mbox{}+C_{F}T_{R}
\left[-\frac{160}{9}S_{1}(J)+\frac{32}{3}S_{2}(J)+\frac{4}{3}
+\frac{16}{9}\frac{(11J^{2}+5J-3)}{J^{2}(J+1)^{2}}\right]\nn\eeq

\beq\label{DIO} \gamma^{(1)}_{qq}(J)&=& \gamma^(1)_{NS}(J) - 16
C_{F}T_{R}\left[ \frac{
(5J^{5}+32J^{4}+49J^{3}+38J^{2}+28J+8)}{(J-1)J^{3}
(J+1)^{3}(J+2)^{2}}\right]\\ & & \nn\\& & \nn\eeq

\beq\label{TRIA} \gamma^{(1)}_{qg}(J)&=& -8
C_{F}T_{R}\left[\frac{(4+8J+15J^{2}+26J^{3}+11J^{4})}{J^{3}(J+1)^{3}(J+2)}
-\frac{4 S_{1}(J)}{J^{2}}\right.\\& &\left.\mbox{}
+\frac{(2+J+J^{2})(5+2S_{1}^{2}(J)-2S_{2}(J))}{J(J+1)(J+2)}\right]\nn\\&
&\mbox{}- 8
C_{A}T_{R}\left[\frac{2(16+64J+104J^{2}+128J^{3}+85J^{4}+36J^{5}+25J^{6}+
15J^{7}+6J^{8}+J^{9})}{(J-1)J^{3}(J+1)^{3}(J+2)^{3}}\right.\nn\\&
&\left. \mbox{} +
\frac{8(3+2J)S_{1}(J)}{(J+1)^{2}(J+2)^{2}}+\frac{(2+J+J^{2})(-2S_{1}^{2}(J)
+2S_{2}(J)-2S_{2}^{\prime}(J))}{J(J+1)(J+2)}\right]\nn\eeq

\vskip1truecm

\beq\label{TESERA} \gamma^{(1)}_{gq}(J)&=&- \frac{32}{3}
C_{F}T_{R}\left[\frac{1}{(J+1)^{2}}+\frac{(2+J+J^{2})(-8/3+S_{1}(J)}
{(J-1)J(J+1)}\right]\\ & &\nn \\
 & &\mbox{}-
4C_{F}^{2}\left[-\frac{(-4-12J-J^{2}+28J^{3}+43J^{4}+30J^{5}+12J^{6})}
{(J-1)J^{3}(J+1)^{3}}-\frac{4S_{1}(J)}{(J+1)^{2}}\right.\nn\\& &
\nn\\& &\left.\mbox{}+
\frac{(2+J+J^{2})(10S_{1}(J)-2S_{1}^{2}(J)-2S_{2}(J))}{(J-1)J(J+1)}\right]
\nn\\& & \nn\\ & &\mbox{} - 8C_{A}C_{F}\left[
\frac{(144+432J-152J^{2}-1304J^{3}-1031J^{4}+695J^{5})}{9(J-1)^{2}J^{3}
(J+1)^{3}(J+2)^{3}} \right.\nn\\& &\nn \\&
&\left.\mbox{}+\frac{(+1678J^{6}+1400J^{7}+621J^{8}+109J^{9})}{9(J-1)^{2}
J^{3}(J+1)^{3}(J+2)^{2}} \right. \nn\\& & \nn\\& &\left.\mbox{}
-\frac{(-12-22J+41J^{2}+17J^{4})S_{1}(J)}{3(J-1)^{2}J^{2}(J+1)}+
\frac{(2+J+J^{2})(S_{1}^{2}(J)+S_{2}(J)-S_{2}^{\prime}(J)}
{(J-1)J(J+1)}\right]\nn\eeq

\vskip1truecm

\beq\label{PENTE} \gamma^{(1)}_{gg}(J) &=& C_{F}T_{R}\left[8 +
\frac{16(-4-4J-5J^{2}-10J^{3}+J^{4}+4J^{5}+2J^{6})}
{(J-1)J^{3}(J+1)^{3}(J+2)}\right]\\& &\nn\\ & &
\mbox{}+C_{A}T_{R}\left[\frac{32}{3}
+\frac{16(12+56J+94J^{2}+76J^{3}+38J^{4})}{9(J-1)J^{2}(J+1)^{2}(J+2)}-
\frac{160 S_{1}(J)}{9}\right]\nn\\& & \nn\\& &\mbox{}+C_{A}^{2}
\left[ -\frac{4(576+1488J+
560J^{2}-1632J^{3}-2344J^{4}+1567J^{5})}{9(J-1)^{2}J^{3}(J+1)^{3}(J+2)^{3}}
\right.\nn \\&& \mbox{} \left.+\frac{(6098J^{3}+6040J^{4}
+2742J^{5}+457J^{6})}{9(J-1)^{2}(J+1)^{3}(J+2)^{3}}\right.\nn\\& &
\nn\\ & &\mbox{}-\frac{64}{3} + \frac{536}{9}S_{1}(J) +
\frac{64(-2-2J+7J^{2}+8J^{3}+5J^{4}+2J^{5})S_{1}(J)}
{(J-1)^{2}J^{2}(J+1)^{2}(J+2)^{2}}\nn\\& &\nn \\& &\left.\mbox{}+
\frac{32(1+J+J^{2})S_{2}^{\prime}(J)}{(J-1)J(J+1)J+2)}-
16S_{1}(J)S_{2}^{\prime}(J) + 32\tilde{S}(J) -4
S_{3}^{\prime}(J)\right]\nn\eeq

\vskip1truecm

 \be\label{EKSI} \gamma_{qq}^{(0)}(J)= 2 C_{F}\left[
4S_{1}(J)-3-\frac{2}{J(J+1)}\right]\ee

\vskip0.8cm

\be\label{EPTA} \gamma_{qg}^{(0)}(J)=-
\frac{8T_{R}(J^{2}+J+1)}{J(J+1)(J+2)}\ee

 \vskip0.8cm
 \be\label{OKTO} \gamma^{(0)}_{gq}(J)= -\frac{4
C_{F}(2+J+J^{2})}{(J-1)J(J+1)}\ee

\vskip0.8cm

 \beq\label{ENIA} \gamma_{gg}^{(0)}(J)&=& \frac{8}{3} T_{R} +
2C_{A} \left[ -\frac{11}{3}- \frac{4}{J(J-1)}
-\frac{4}{(J+1)(J+2)} +4 S_{1}(J) \right] \eeq \vskip0.8cm





\beq && S_{n}(J)= \sum_{k=1}^{J} \frac{1}{k^{n}}\, , ~~~~
S^{\prime}_{n}(J)= 2^{n-1}\sum_{k=1}^{J} \frac{1 +
(-1)^{k}}{k^{n}}\nonumber \\&&   ~~~~~ \tilde{S}(J)= \sum_{k}^{J}
\frac{(-1)^{k}S_{1}(k)}{k^{2}} \eeq

Here $C_{F}$,$ C_{A}$ are the Casimirs for the fermions and gauge
bosons whereas $T_{R}$ is one half the number of fermions. To
facilitate the reader, we present the large $J$ behaviour of some
harmonic functions. In particular, from the above definitions it
follows that
\beq
S_n(J)-S_n'(J)\to 0~~~\mbox{for}~~~J\to \infty
\eeq
Moreover, we have
\begin{eqnarray}
S_1(J)\sim \log(J)+C+{\cal{O}}(J^{-1})\, , &&S_2(J)\sim\frac{\pi^2}{6}+{\cal{O}}(J^{-1})\\
S_3(J)=\zeta(3)+{\cal{O}}(J^{-2})\, ,&& \tilde{S}(J)\sim-\frac{5}{8}\zeta(3)+
\frac{(-1)^J}{2}\frac{\log{J}}{J^2}+{\cal{O}}(J^{-2})\, .
\end{eqnarray}



\setcounter{equation}{0}
\section{{\cal N}=1 SUSY Yang-Mills Anomalous Dimensions of Wilson
Operators}

 In the following we review the results of \cite{Flo3} and proceed to
 obtain the explicit form of the anomalous dimensions
for the ${\cal N}=1$ supersymmetric case at the two loop level
from the already known results in QCD. To this end we put the
fermions in the adjoint representation and consider only the ones
of the Majorana type. This can be obtained directly from the
non-susy results for the special case of $C_F=C_A=2 T_R=N$.
 However, beyond one-loop, the dimensional renormalization scheme ${\rm
\bar{MS}}$, in which  the QCD results mentioned above are
obtained, breaks supersymmetry as well as does the covariant gauge
fixing. The latter is solved by  calculating anomalous dimensions
of gauge invariant observables like the Wilson operators. In order
to preserve supersymmetry, we have to use the dimensional
reduction scheme (DR) in which the momentum integrations are done
in $4-\epsilon$ dimensions and the spin-index
 algebra is performed in $4$ dimensions. Instead  of repeating the long
two-loop calculations, there is a way to pass from  ${\rm \bar{MS}}$
 to DR if we calculate the one-loop finite part of Wilson operators
in both schemes and the relation between the ${\cal {N}}=1$
 YM gauge couplings  between the two schemes at two loops
\beq
{\alpha_{s}}_{DR}={\alpha_s}_{\bar{MS}}+{1\over 3}
{\alpha_s}_{\bar{MS}}^2+...
\eeq

For the singlet anomalous dimensions, the relevant transformation
rule between the two schemes at two-loops is \cite{Flo3}
\beq
\gamma^{(1)}_{DR}+b_0 O^{(0)}_{DR}+[\gamma^{(0)}_{DR},O^{(0)}_{DR}]=
\gamma^{(1)}_{\bar{MS}}+b_0 O^{(0)}_{\bar{MS}}+[\gamma^{(0)}_{\bar{MS}},
O^{(0)}_{\bar{MS}}]-{1\over 3} \gamma^{(0)}_{\bar{MS}} \label{AF}
\eeq
where the two-by-two matrix of the finite parts of Wilson operators
is defined as
\beq
O=\left(\begin{array}{cc}
O_{qq}&O_{qg}\\
O_{gq}&O_{gg}
\end{array}\right)
\eeq \beq O=\alpha_s O^{(0)}+\alpha_s^2O^{(1)}+... \eeq

By employing eq.$(5.2)$ and the two loop results from the Appendix
we find the ${\cal N}=1$ supersymmetric singlet anomalous
dimensions. In the following we will omit the overall factor
$C_{A}^{2}$  as well as the $ -\frac{1}{3}\gamma^{(0)}_{\rm
\bar{MS}}$ contribution. The $\gamma^{(1)}$'s which are portrayed
below are calculated in the DR scheme \vskip1truecm
\beq\label{ENA} \gamma_{qq}^{(1s)}(J)&=& -14 + \frac{8\,\left( 2 +
J + J^2 \right) }
   {J\,{\left( 1 + J \right) }^2\,\left( 2 + J \right) }
- \frac{4\,\left( 18 + 39\,J + 142\,J^2 + 290\,J^3 + 151\,J^4 \right) }
   {9\,J^3\,{\left( 1 + J
 \right) }^3}
\nonumber \\
&& -
  \frac{8\,\left( 8 + 28\,J + 38\,J^2 + 49\,J^3 + 32\,J^4 + 5\,J^5 \right) }
   {\left( -1 + J \right) \,J^3\,{\left( 1 + J \right) }^3\,{\left( 2 + J
\right) }^2}+
  \,\left(  \frac{2}{J\,\left( 1 + J \right) } \right) \nonumber \\
&&  +
  \frac{8\,\left( -3 + 11\,J^2+18 S_1(J)+J(5+36 S_1(J) \right) }{9 J^2\,
{\left( 1 + J \right) }^2}  +
  8\,\left(  \frac{1}{J\,\left( 1 + J \right) }   -
2\,S_1(N) \right)
      S_2'(J)\nonumber \\
&& + \frac{152\,S_1(J)}{3}+
  32\,\tilde{S}(J) - 4\,S_3'(J)\nonumber \\& & \nonumber \\& & \nonumber \\
\gamma_{qg}^{(1s)}(J)&=&\frac{4\left( 2 + J + J^2 \right) \, }
   {J\,\left( 1 + J \right) \,\left( 2 + J \right) }
\left(- {10\over 3} - \frac{2}{J}  + \frac{6}{1 + J} - \frac{4}{2 +
J} +  2\,S_2'(J) \right) \nonumber \\
&&-
  \frac{4\left(4 + 8\,J + 15\,J^2 + 26\,J^3 + 11\,J^4\right)}
      {J^3\,{\left( 1 + J \right) }^3\,\left( 2 + J \right) } +
     \frac{16\,S_1(J)}{J^2}- \frac{32\,\left( 3 + 2\,J \right) \,S_1(J)}
      {{\left( 1 + J \right) }^2\,{\left( 2 + J \right) }^2} \nonumber \\
&&  -
   \frac{8\,\left( 16 + 64\,J + 104\,J^2 + 128\,J^3 + 85\,J^4 +
36\,J^5 +
25\,J^6 +
          15\,J^7 + 6\,J^8 + J^9 \right) }{\left( -1 + J \right) \,J^3\,
\left( 1 + J \right)^3\,
        \left( 2 + J \right)^3}\nonumber\\
& & \nonumber \\ \gamma_{gq}^{(1s)}(J)&=&- \frac{8\,\left( 2 + J +
J^2 \right) \,\left( -16 + 51\,S_1(J) - 9\,S_2'(J) \right)
}{9\,J\,\left( -1 + J^2 \right) } +\frac{16\,\left( -1 + 3\,S_1(J)
\right) }{3\,{\left( 1 + J \right) }^2}\nonumber \\ &&+
\frac{4\,\left( -4 - 12\,J - J^2 + 28\,J^3 + 43\,J^4 + 30\,J^5 +
12\,J^6 \right) }
  {\left( -1 + J \right) \,J^3\,{\left( 1 + J \right) }^3}
\nonumber \\
&&-\frac{8\,\left( 144 \!+\! 432 J\! -\! 152J^2\! \!- 1304J^3\! -\!
 1031J^4 \!+\! 695J^5\!
+\! 1678J^6\! +\!
      1400J^7 \!+\! 621J^8\! +\! 109J^9 \right) }{9\,J^3\,
{\left( 1 + J \right) }^3\,
    {\left( -2 + J + J^2 \right) }^2}\nonumber \\
&&+
\frac{8\,\left( -12 - 22\,J + 41\,J^2 + 17\,J^4 \right) \,
S_1(J)}
  {3\,{\left( -1 + J \right) }^2\,J^2\,\left( 1 + J \right) }\nonumber \\
& & \nonumber\\& & \nonumber \\ \gamma_{gg}^{(1s)}(J)&=&
  \frac{4\left( -\!576\!-\! 1488J \!-\! 560J^2 \!+\! 1632J^3 \!+\!
 2344J^4 \!-\! 1567J^5\! -\!
6098J^6 \!- \!
       6040J^7\! -\! 2742J^8 \!- \!457J^9 \right) }{9\,
{\left( -1 + J \right) }^2J^3\,
     {\left( 1 + J \right) }^3{\left( 2 + J \right) }^3} \nonumber \\
&& + \frac{16\,\left( -1 +J + J^2 \right) }
   {J\,{\left( 1 + J \right) }^2\,\left( 2 + J \right) }
   + \frac{8\,\left( -4 - 4J - 5\,J^2 - 10\,J^3 + J^4 + 4\,J^5 + 2
\,J^6 \right) }
      {\left( -1 + J \right) \,J^3\,{\left( 1 + J \right) }^3\,\left( 2 + J
\right) } +\nonumber \\
&&
  + \frac{8\,\left( 12 + 56\,J + 94\,J^2 + 76\,J^3 + 38
\,J^4 \right) }
      {9\,\left( -1 + J \right) \,J^2\,{\left( 1 + J \right) }^2\,
\left( 2 + J \right) } +
  \frac{32\,\left( 1 + J + J^2 \right) \,S_2'(J)}
   {\left( -1 + J \right) \,J\,\left( 1 + J \right) \,\left( 2 + J
\right) }
 \nonumber \\
&&+
  \frac{64\,\left( -2 - 2\,J + 7\,J^2 + 8\,J^3 + 5\,J^4 + 2\,J^5 \right)
\,S_1(J)}
   {{\left( -1 + J \right) }^2\,J^2\,{\left( 1 + J \right) }^2\,
{\left( 2 + J \right) }^2}
\nonumber \\
&&  -14+ \frac{456\,
S_1(J)}{9} -
  16\,S_1(J)\,S_2'(J) + 32\,
\tilde{S}(J) -
  4\,S_3'(J)) \\& & \nonumber \\& & \nonumber
\eeq

It is easy to check that the $\gamma^{(1)}(J)$'s satisfy as well
the Dokshitzer relation of eq.($4.1$).


 In this section we discuss the large spin behavior of the anomalous
dimensions of Wilson operators for supersymmetric and
non-supersymmetric Yang-Mills theories. This question is relevant
for the gauge theory-string duality as we have discussed above. It
is also of phenomenological interest as the scaling violations
become more prominent in the kinematic regime of $x_B\to 1$
\cite{Gross, GLY}. The Feynman rules for the Wilson operators
dictate that the diagrams with gluon lines coming out from the
operator vertex will contribute terms which behave like $
\lambda^{k}\,\,(\log J)^{2k-1}$ \cite{Gross} and sub-dominant
terms at k-loop order.

In the nonsupersymmetric case the asymptotic over all $log J$
behavior comes about after successive miraculous cancellations of
all the $(log J)^2$ and $(log J)^3$ terms at two loops inside the
gauge invariant classes of diagrams.

This cancelation is a
consequence of Ward indentities for the gluon-quark-quark vertex with
an insertion of the quark operator $O^N_q$. As an illustration the leading
behaviour of the quark-quark diagramms of Fig.1 is
(we do not include explicitly
the $\log J$ terms of these diagrams)
\beq
&&{\rm diag.~~ A}: ~~~~~~~4(\log J)^2 C_F^2 \nonumber \\
&&{\rm diag. ~~B}: ~~~~~~~-{2\over 3}(\log J)^3 C_F C_A\nonumber \\
&&{\rm diag. ~~C}: ~~~~~~~-4 (\log J)^2 (C_F^2-{1\over 2} C_F C_A)\nonumber \\
&&{\rm diag. ~~D}: ~~~~~~~\left(-{2\over 3}(\log J)^3 +
2 (\log J)^2\right) C_F C_A\nonumber \\
&&{\rm diag. ~~E}: ~~~~~~~\left({4\over 3}(\log J)^3 -
4 (\log J)^2\right) C_F C_A
\eeq

At this point it is important to observe that  supersymmetry ($
C_{F}=C_{A}$) does not improve the $\log J$  behavior of the above
diagrams. From the previous section on the other hand  there
appears some cancellations in the subleading terms of
$\gamma_{qg}^{1s}$(especially those which behave like $(\log
J)^{2}/J$ for large $J$).

We exhibit below the large $J$ behaviour and large $N$ of both the
nonsupersymmetric and suppersymmetric anomalous dimensions at one
and two loops. We absorb the $N$ factors of the Casimirs in the 't
Hooft coupling. In this limit we typically expand the anomalous
dimensions as follows: \beq \gamma^{J}(\lambda)= {\lambda\over 4
\pi}\gamma^{(0)}(J)+ ({\lambda\over 4 \pi})^2
\gamma^{(1)}(J)+\cdots \eeq

In particular the non-supersymmetric asymptotic behaviour
 ($C_A=N$, $C_F=\frac{N^2-1}{2 N}$ and $T_R=
\frac{n_F}{2}$ for large $N$ gives ($C_A=2 C_F=N, T_R=0$),  is
 for one and two loops, respectively:
\beq
\gamma_{qq}^{(0)}(J)&\sim& 4 \log J + 4 \gamma-3\nonumber \\
\gamma_{gg}^{(0)}(J)&\sim&2\gamma_{qq}^{(0)}(J)\nonumber \\
\gamma_{qq}^{(1)}(J)&\sim& {4\over 9}(67-3 \pi^2) \log J-{1\over 36}
(129 +52 \pi^2
+16 \gamma(3\pi^2-67))\nonumber \\
\gamma_{gg}^{(1)}(J)&\sim&2\gamma_{qq}(J)
\eeq
where $\gamma$ is the Euler-Masceroni constant.

In the ${\cal N}=1$ supersymmetric case ($C_A=C_F=2T_R=N$) the
asymptotic behaviour in the $DR$ scheme is given by: \beq
\gamma_{qq}^{(0)}(J)&\sim&8 \log J + 8 \gamma-6\nonumber \\
\gamma_{gg}^{(0)}(J)&\sim&\gamma_{qq}^{(0)}(J)\nonumber \\
\gamma_{qq}^{(1)}(J)&\sim&{8\over 3}(19-\pi^2)\log J-{2\over 3}
(21+36 \zeta(3)+4 \gamma(\pi^2-19))\nonumber \\
\gamma_{gg}^{(1)}(J)&\sim&\gamma_{qq}^{(1)}(J) \eeq

As far as the off-diagonal elements is concerned the one- and
two-loop singlet anomalous dimensions tend to zero in  both the
supersymmetric and non-supersymmetric case. At this point we
remark that there is important piece of literature for the
resummation methods of the leading behaviour at one and two loops
of the structure functions near $x_{B}\rightarrow 1$ \cite{Lip}.
We would like to draw attention to the study of this limit through
the cusp singularities of the Wilson loop\cite{KM}

\section*{Conclusions}

We would like to summarize our work by the following observations.
The $ (\log J)$ behaviour of both the singlet and non-singlet
anomalous dimensions for QCD was known for a long
time\cite{GW,GLY}. It constitutes a widespread belief that it also
holds true to all orders in perturbation theory. This amounts to a
$ (1-x)^{-1}$ behaviour of the GLAP splitting function as
$x\rightarrow 1$. In the present work we proved that this holds
true as well for the ${\cal N}=1$ supersymmetric Yang-Mills
theory. This is not obvious because by passing in the
supersymmetric case from the $\bar{MS}$ to the $DR$ scheme one
must include the one-loop finite part of the Wilson operators.
These matrix elements contain $(\log J)^2$ terms for large $J$ but
their coefficients are the same in the two-schemes. From relation
eq.(\ref{AF}) it follows that it is only the difference between
the matrix elements for the two schemes participates and so $(\log
J)^2$ terms cancel. On the string theory side in $AdS_{5}$ an
identical claim holds but for the strong coupling t'Hooft limit
$\frac{\alpha_{s}}{4\pi} N\rightarrow \infty$. We suspect the
presence of a geometrical reason as a consequence of which long
strings which touch the horizon violate by simple logarithmic
power the Energy-Spin relation.

We hope that soon three loop results will become available for QCD
and the transition from the ${\cal N}=0$ to ${\cal N}=1,2,4 $
cases will be possible. The expected existence of dualities for
the ${\cal N} = 4$ case between the strong and weak t'Hooft
coupling will hopefully as well become available for anomalous
dimensions \cite{Lip}. Thus the explicit demonstration, to all
orders, of the validity of the GKP conjecture on both sides of the
geometric duality appears to be both interesting and challenging a
problem.

 \vskip.5cm

\paragraph{Acknowledgements:}
We would like to thank I. Antoniadis and C. Kounnas for
discussions. This work is partially supported by European RTN networks
HPRN-CT-2000-00122 and HPRN-CT-2000-00131.
\vskip.5cm

%

\newpage

\begin{figure}[!h]
\centerline{\hbox{\psfig{figure=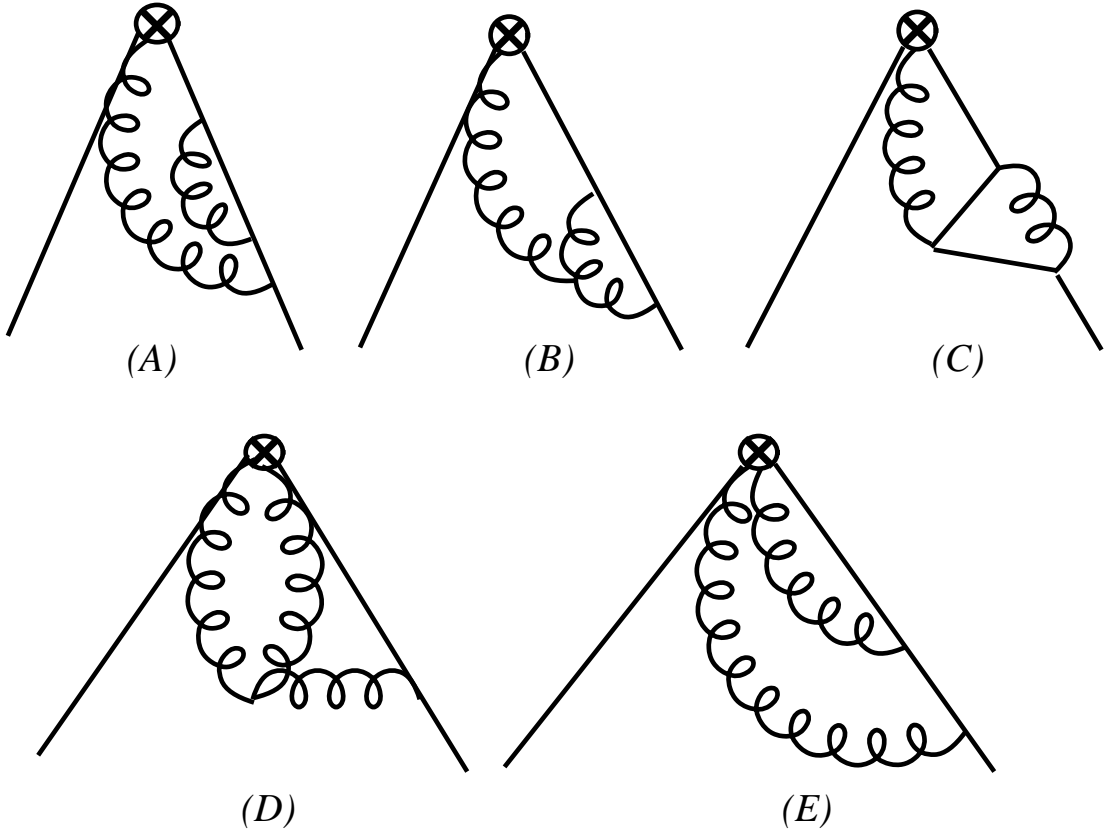,height=15cm}}}
\caption{{\bf Two-loop quark-quark diagramms  with quadratic or cubic
in $\log J$ leading behavior at large spin $J$}
{\footnotesize }}
\label{f6}
\end{figure}
\newpage


\newpage

\end{document}